\documentclass[aps,twocolumn,floatfix]{revtex4}

\usepackage{amssymb,amsmath}
\usepackage{graphicx}
\usepackage{subfigure}

\begin{document}

\title{Universality of the rainfall phenomenon}
%

%
%





\author{M. Ignaccolo$^{1}$}
\author{C. De Michele$^{2}$}
\author{S. Bianco$^{1}$}

\affiliation{$^{1}$University of North Texas, Center for Nonlinear Science, Department of Physics, Denton, TX, United States of America.}

\affiliation{$^{2}$Politecnico di Milano, Department of Hydraulic, Environmental, Roads and Surveying Engineering, Milano, Italy.}

\begin{abstract}
We show that the universal properties of the rainfall phenomenon are the scaling properties 
of the probability density function of inter-drop intervals during quiescent periods, time intervals of sparse precipitation,
 and the universal shape of the probability density function of drop diameters during non-quiescent periods, time intervals of 
active precipitation. Our results indicate that the continuous flux-like vision of rainfall based on quantities such as the 
rain duration, rain intensity and drought duration is ineffective in detecting the universality of the phenomenon. A comprehensive 
understanding of rainfall behavior must rest on the acknowledgment of its discrete drop-like nature.
\end{abstract}

\maketitle

\section{Introduction}
Rainfall is a discrete drop-like phenomenon that has often been described as a continuous 
flux-like phenomenon. The most common instrument used to measure the rain, the pluviometer, collects the water volume fallen 
through a given area per unit of time. The use of pluviometers and the importance of knowing intensity and duration of the rainfall phenomenon 
has lead to a description based on flux-like quantities such as the rain duration, rain intensity and drought 
duration \citep{PSE70}, even when radar measurement are used to infer the 
precipitable volume of rain \citep{PHC01}. A flux-like view of rainfall is also central to the random cascade formalism used to 
describe rainfall pattern both in time and space \citep[e.g.,][]{LS85,LS06,GW90,MH97}.
From a drop-like perspective, a considerable amount of work \citep[e.g.,][]{MP48,JG78,AU82,FL86} has been dedicated to the study the properties 
of raindrop spectra: the number of drops per diameter millimeter interval per cubic meter of air. Double stochastic Poisson processes have 
been used to describe the variability in the drop counts per unit interval \citep[e.g.,][]{SM93,JK00}. Only recently \citep{LG98, LG06}
an extensive study has been done of the properties of  the sequences of inter drop time intervals and drop diameters as measured on the ground 
by disdrometers. Due to its importance for many aspects of human life, the rainfall phenomenon has been widely investigated. However, few works 
\citep[e.g:][]{WT05} are explicitly dedicated to discuss the universal properties of the rainfall phenomenon. No entries were found (using common 
literature search engines) for works containing in their titles both the words ``universality'' (or universal) and ``rainfall'' (or rain).  
So, are there any properties that a rain shower in New York and one in Rome have in common?

In this letter, using data from the Joss Waldvogel impact disdrometer RD-69 located at Chilbolton (UK), we provide evidences that 
the universal properties of rainfall phenomenon lie in the properties of drop-like quantities such as the inter drop time interval and the 
drop diameter. We show that 1) the flux-like view of the rainfall phenomenon is not adequate to capture its universal features. 2) The temporal 
variability of rain can be described in terms of quiescent periods, periods of sparse precipitation characterized by a small drop diameter 
average, and non quiescent periods, periods of active precipitation characterized by a large drop diameter average diameter. Moreover, the 
average and variance of the sequence of drop diameters are not stationary. 3) The probability density function of the inter drop time 
intervals $\tau$ has an universal feature: a power law regime in the region \mbox{$\tau$$\gtrsim$$1$min} and \mbox{$\tau$$\lesssim$$1$h}. 
Inter drop time intervals in this range belong to quiescent periods. Finally, an universal shape for the probability density function of 
drop diameters during non-quiescent periods emerges upon removal of the non stationarity of the sequence of drop diameters. 
\section{Flux-like view of a Drop-like phenomenon}
\begin{figure} [h]
\includegraphics[angle=-90,width=1.0\linewidth]{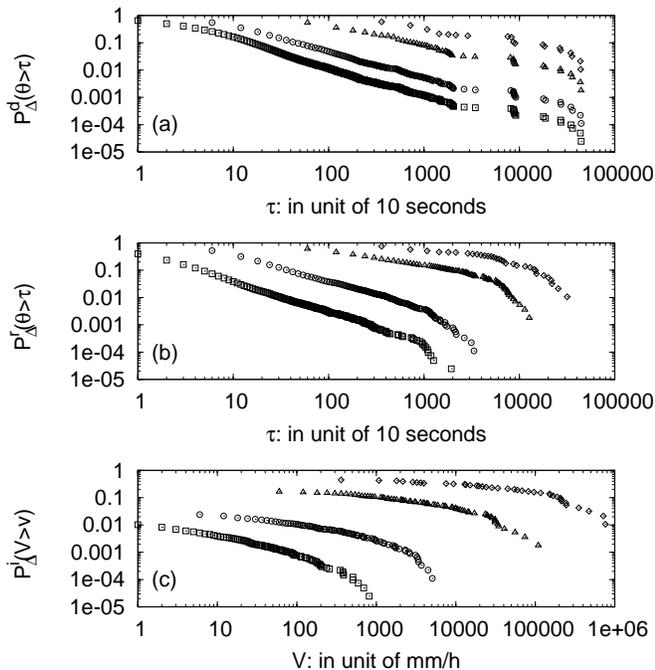}
\caption{Panel (a): Log-log plot of the probability $P^{\textrm{d}}_{\Delta}(\theta$$>$$\tau)$ of having a drought duration larger than $\tau$ 
during the interval of time from January 24th to May 11th 2004. Different symbols indicate different 
values on the integration time intervals: \mbox{$\Delta$$=$$10$} seconds (white squares),  \mbox{$\Delta$$=$$1$} minute (white circles), \mbox{$\Delta$$=$$10$} 
minutes (white triangles), \mbox{$\Delta$$=$$1$} hour (white diamonds). Panel (b): Log-log plot of the probability $P^{\textrm{r}}_{\Delta}(\theta$$>$$\tau)$ of having 
a rain duration larger than $\tau$ during the same interval of time, and for the same the integration time intervals of 
panel (a). Panel (c): The probability $P^{\textrm{i}}_{\Delta}(V$$>$$v)$ of having a rain duration larger than $v$ during the same interval of time, and for 
the same the integration time intervals of panel (a).}
\label{figure1}
\end{figure}

The flux-like view of rainfall is that of a ON-OFF process. The rainfall time series is an alternating sequence of consecutive time intervals of 
duration \mbox{$\Delta$$>$$0$}, the integration time of the instrument used to monitor the precipitation, with (ON) or without (OFF) 
detectable precipitation. The relevant quantities \citep[e.g.,][]{PHC01} are the duration of ON and OFF periods (rain duration and drought duration) 
and the volume of rain fallen during consecutive OFF periods (rain intensity). An integration time \mbox{$\Delta$$>$$0$} causes all drop time intervals 
of duration \mbox{$\tau$$<$$\Delta$} to be lost (equivalent to be detected as a drought of null duration), and all inter drop time intervals of 
duration \mbox{$\tau$$>$$\Delta$} to be detected as drought of duration $\left[\tau / \Delta \right]$$-$$1$ or $\left[\tau / \Delta \right]$ 
($\left[.\right]$ indicates the integer part). Thus, in the limit \mbox{$\tau$$\gg$$\Delta$} 
\begin{equation}\label{niceequation} 
P^{\textrm{d}}_{\Delta}(\tau) \propto \Delta\psi(\tau),
\end{equation}
where $P^{\textrm{d}}_{\Delta}(\tau)$ is the distribution of drought durations and $\psi(\tau)$ is probability density function of inter drop 
time intervals. \mbox{Eq.~(\ref{niceequation})} shows that distributions of drought durations relative to different integration times $\Delta$ 
will all have the same features at large times (\mbox{$\tau$$\gg$$\Delta$}) as confirmed by panel (a) of \mbox{Fig.~\ref{figure1}}. This property is lost 
for the distributions of rain durations $P^{\textrm{r}}_{\Delta}(\tau)$, panel (b) of \mbox{Fig.~\ref{figure1}}, and rain intensities 
$P^{\textrm{i}}_{\Delta}(v)$, panel (c) of \mbox{Fig.~\ref{figure1}}. As the integration time $\Delta$ increases a larger amount of inter drop time 
intervals are lost and rain durations that were separated by a drought duration are now detected as a longer rain period. This effect produces 
larger rain durations and, together with the temporal ordering of the sequence of drop diameters, larger rain intensities.
 
The properties of the sequences of inter drop time intervals and drop diameters do not straightforward translate into those of 
the distributions of rain durations ($P^{\textrm{r}}_{\Delta}(\tau)$) and intensities ($P^{\textrm{i}}_{\Delta}(v)$) \citep[e.g.,][]{S86}. Thus, these quantities 
are not valuable proxies to investigate the universal properties of the rainfall phenomenon.
\section{Temporal Variability}\label{tempvar}
Our data are from a Joss Waldvogel impact disdrometer RD-69 with a time integration \mbox{$\Delta$$=$$10$s}, and 
127 different diameter classes (from 0.2998 mm to 4.9952 mm). 
Thus, we can record neither all the inter drop time intervals less than 10s, nor the exact arrival 
ordering of drops. However, some properties of the temporal variability of the rainfall phenomenon 
can be inferred.
\subsection{Quiescence}\label{quies}
\begin{figure} [h]
\includegraphics[angle=-90,width=1.0\linewidth]{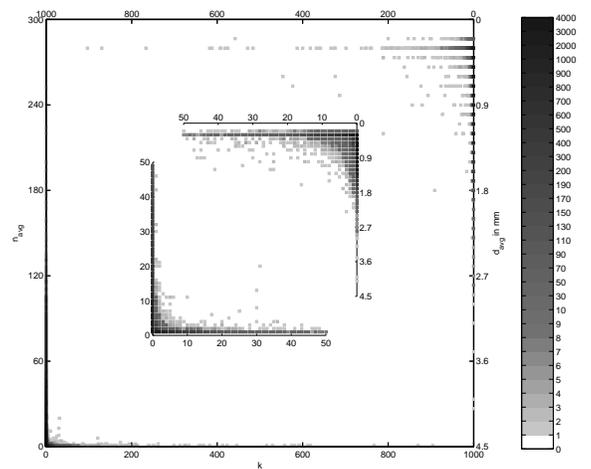}
\caption{The logarithm of the density $\rho$($k$$,$$n_{avg}$) of the couples ($k$$,$$n_{avg}$) in the $k$$n_{avg}$ plane (bottom x-axis and left y-axis).
The logarithm of the density $\rho$($k$$,$$d_{avg}$) of the couples ($k$$,$$d_{avg}$) in the $k$$d_{avg}$ plane (top x-axis and right y-axis). 
The inner plots show the logarithm of the densities $\rho$($k$$,$$n_{avg}$) and $\rho$($k$$,$$d_{avg}$) in the region close to respective origins. 
Both densities are obtained using the data relative to the time interval from January 24th to May 11th 2004.}
\label{figure2}
\end{figure}

Quiescence is a  way of describing the temporal variability of the rainfall phenomenon based on the relationship between inter drop time 
intervals and drop diameters. We divide the rainfall time series in time intervals of length $\delta$. A couple of consecutive time intervals 
of detectable precipitation may be separated by $k$$>$$0$ consecutive droughts (empty intervals) or be adjacent. In this case we say that the 
couple is separated by $k$$=$$0$ consecutive droughts. Finally, for each couple we evaluate the average number of drops $n_{avg}$, and the 
average drop diameter $d_{avg}$. \mbox{Fig.~\ref{figure2}} shows the relationship between the couple average number of drops, the couple average drop 
diameter and the number of consecutive droughts in between the couple for \mbox{$\delta$$=$$10$s}. We see the tendency for a number of consecutive droughts 
\mbox{$k$$\geq$$1$ (interdrop time intervals $\tau$$\geq$$10$s)} to separate couples with a small average number of drops \mbox{($\lesssim$$5$)} and a small 
average drop diameter \mbox{($\lesssim$$0.6$mm)}. To quantify this tendency, we introduce the concept of quiescence of order \mbox{$(\delta,m,n)$}. 
A couple of consecutive time intervals of length $\delta$ with detectable precipitation and separated by $k$ consecutive droughts is a quiescent couple
of order \mbox{$(\delta,m,n)$}, if:
\begin{equation}
\label{quiescence}
k \geq m \;\;\;\;\;\; \textrm{or} \;\;\;\;\;\; k<m \;\; \textrm{but} \;\; n_{\textrm{avg}}\leq n,
\end{equation}
where $n_{\textrm{avg}}$ is the couple average number of drops. 

Quiescent (non-quiescent) periods are rainfall periods occupied by consecutive quiescent (non-quiescent) couples.
During quiescent periods, large inter drop intervals (\mbox{$\tau$$\geq$$m$$\delta$)} are preceded and followed by few \mbox{($n_{\textrm{avg}}$$\leq$$n$)} 
drops of small diameter (\mbox{Fig.~\ref{figure2}}). Non-quiescent or active periods are characterized by small inter drop time 
intervals ($\tau$$<$$m$$\delta$) separating drops with a larger range of diameters (\mbox{Fig.~\ref{figure2}}). Non quiescent periods are responsible 
for the bulk of precipitation: for a quiescence of order \mbox{$(\delta$$=$$10$s$,m$=$$1$,n$=$$5$)$} \mbox{$>$$95$\%} of the precipitated volume of 
water belongs to non quiescent periods. Some care is necessary in choosing the duration $\delta$ of the time intervals dividing the rainfall time 
series: a $\delta$ too large \mbox{($\gtrsim$$10$ minutes)} will result in a mixing quiescent with non-quiescent periods, a $\delta$ too small 
\mbox{($\lesssim$$1$ second)} will result in too many time integration intervals with just one drop. Two quiescence of order  \mbox{$(\delta,m,n)$} 
and \mbox{$(\delta^{\prime},m^{\prime},n^{\prime})$} are ``equivalent'' if \mbox{$m^{\prime}$$=$$m$$(\delta/\delta^{\prime})$} and 
\mbox{$n^{\prime}$$=$$n$$(\delta/\delta^{\prime})$}.
\subsection{Drop diameter variability}\label{ddvar}
\cite{JG78} introduce the concept of averaged ``instant'' shape to characterized the variability of raindrop spectra and their departure from 
the exponential form observed by \cite{MP48}. \cite{PK97} show that the probability density function $\eta(d)$ of drop diameters may change according to 
the portion (e.g.: dissipative edge, cloud base) or the type of storm (e.g.: orographic, non orographic) observed by a disdrometer. Here, we show the 
temporal variability of the sequence of drop diameter is characterized by a moving average and a moving variance. 
Fig.~\ref{figure2} indicates that the sequence of drop diameters does not have a constant average, as quiescent periods have a lower average diameter 
than non-quiescent periods. A closer examination of the sequence of drop diameters reveals that its average together with its variance are not 
stationary. Support for this thesis comes also from the results of \cite{LG06}. They report for the autocorrelation function 
of the sequence of drop diameters with a slow decay (the autocorrelation function reaches zero at lag \mbox{$\approx$1250}) followed by a long negative 
tail (lag \mbox{$\gtrsim$1250}). The auto correlation function (as confirmed by simulations not reported here for brevity) of sequences of drop diameters 
exponentially and normally distributed with changing intensity around a moving average has the same features of that of \cite{LG06}. 

\section{Universality}\label{univ}
\begin{figure} [h]
\includegraphics[angle=-90,width=1.0\linewidth]{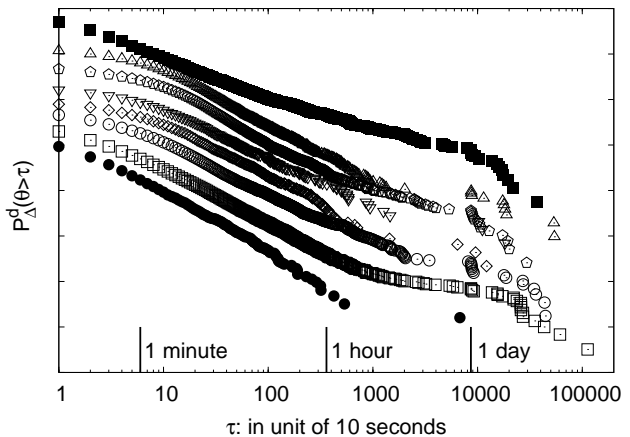}
\caption{Log-log plot of the probability $P^{\textrm{d}}_{\Delta}(\theta$$>$$\tau)$ 
of having a drought of duration $\theta$ larger than $\tau$ as a function of $\tau$ for different 
time intervals of continuous observations at Chilbolton. The period of continuous observation were: from April 1st to November 3rd 2003 (white squares), 
from November 5th 2003 to January 5th 2004 (black squares), from January 8th to January 20th 2004 (black circles), from January 24th to May 11th 2004 
(white circles), from May 14th to July 17th 2004 (white upward triangles), from July 19 th to August 2nd 2004 (white downward triangles), from August 4th to 
August 19 th 2004 (white diamonds), from December 10th 2004 to February 28th 2005 (white pentagons). The curves are shifted for clarity. The ticks on the 
y axis indicates different decades.}
\label{figure3}
\end{figure}

In Fig.~\ref{figure3}, we plot the probability $P^{\textrm{d}}_{\Delta}(\theta$$>$$\tau)$ of having a drought duration larger than $\tau$ 
for several time intervals of continuous observations at Chilbolton over a period of almost 2 years.  
All curves show a power law regime in the region between \mbox{$\tau$$\gtrsim$$1$min} and \mbox{$\tau$$\lesssim$$1$h}. The extensive period of time 
covered by our data, together with observations in other location of Earth's surface \citep{LG98,PHC01}, indicate that  
the power law regime in the region from $1$ minute to $1$ hour of  \mbox{Fig.~\ref{figure3}} 
is an universal property of the probability density function $\psi(\tau)$ of inter drop time 
intervals (Eq.~\ref{niceequation}). This power law regime is a characteristic of quiescent periods:  
all quiescent couples (Eq.~\ref{quiescence}) of order ($\delta$$=$$10$s, $m$$=$$1$, $n$$=$$5$) are separated by inter drop time intervals 
\mbox{$\tau$$\geq$$10$s} (Fig.~\ref{figure2}). Moreover, \mbox{Fig.~\ref{figure3}} suggests that the end of the power law regime at 
\mbox{$\tau$$\approx$$1$h} signals a time scale separation between two different dynamics: the inter storm dynamics where quiescent and non-quiescent 
periods alternate each other, and the dynamics regulating the occurrence of different storms (meteorological dynamics). Thus, 
the probability density function $\psi(\tau)$ of inter drop time intervals can be thought as the sum of three components: 1) $\psi_{\textrm{NQ}}(\tau)$, 
the probability density function of non quiescent periods ($\tau$$\in$$\left[0\textrm{s},\lesssim10\textrm{s}\right]$).
2) $\psi_{\textrm{Q}}(\tau)$, the probability density function of quiescent periods ($\tau$$\in$$\left[\gtrsim0\textrm{s},\lesssim1\textrm{h}\right]$) with a 
power law  regime in the region between \mbox{$\tau$$\gtrsim$$1$min} and \mbox{$\tau$$\lesssim$$1$h}. 3) $\psi_{\textrm{QM}}(\tau)$, the probability 
density function describing the meteorological variability (\mbox{$\tau$$\gtrsim$$1$h}) of the particular location where the measurements are done. 
The index Q in $\psi_{\textrm{QM}}(\tau)$ indicates that all inter drop time intervals $\tau$$\gtrsim$1h belongs to quiescent periods (Eq.~\ref{quiescence} 
and Fig.~\ref{figure2}).

\begin{figure} [h]
\includegraphics[angle=-90,width=1.0\linewidth]{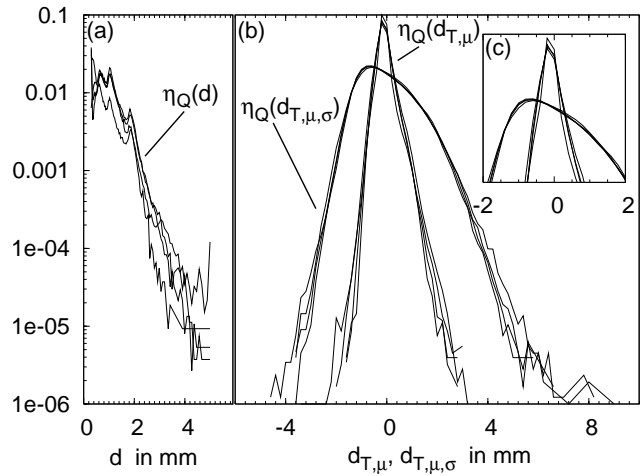}
\caption{Panel (a): The probability density function $\eta_{Q}(d)$ of 
drop diameters during non-quiescent periods for the months of April 2003, 
October 2003, March 2004 and April 2004. Panel (b): The probability density 
function $\eta_{Q}(d_{T,\mu})$ of zero-average drop diameters, and the 
probability density function $\eta_{Q}(d_{T,\mu,\sigma})$ of zero-average 
unitary-variance drop diameters during non-quiescent periods for the months 
of April 2003, October 2003, March 2004 and April 2004. Panel (c): Zoom of 
panel (b) for the values of $d_{T,\mu}$, $d_{T,\mu,\sigma}$ in the interval 
$[$$-2$,$2$$]$, and for values of the probability density in the interval 
$[$$0.01$,$0.1$$]$. The non stationarity of the sequence of drop diameters for quiescence periods 
was removed using time intervals of duration \mbox{$T$$=$$10$}.} 
\label{figure4}
\end{figure}

In panel (a) of Fig.~\ref{figure4}, we plot the probability density functions $\eta_{Q}(d)$ of drop diameters of quiescent periods relative to 
different months of observations. The observed variability is due to the non stationary character of the sequence of drop diameters 
(Sec.~\ref{ddvar}). In fact, if the non stationarity is removed an universal shape for the probability density function emerges. 
We consider non-overlapping time intervals of duration $T$ and remove the average in every time interval. 
Panel (b) of Fig.~\ref{figure4} shows that the probability density function $\eta_{Q}(d_{T,\mu})$ of the zero-average 
drop diameters sequence has a much smaller variability than the probability density function  $\eta_{Q}(d)$ of 
the original sequence (Fig.~\ref{figure4} panel (a)). If together with moving average also the moving variance is 
eliminated (e.g. rescaling to unity the variance in each time interval), the probability 
density functions $\eta_{Q}(d_{T,\mu,\sigma})$ of the zero-average unitary-variance drop diameter sequences 
relative to different months ``collapse'' into a single curve (Fig.~\ref{figure4} panels (b) 
and (c)).  The shape of this curve is not appreciably altered either by the choice of time intervals of different duration $T$ 
(ranging from $10$ seconds to $\approx$$10$ minutes) to remove the non stationarity of the sequence of drop diameters of quiescent periods, either by the use of 
different quiescence orders ($n$ of Eq.~(\ref{quiescence}) ranging from 5 to 20) and of their ``equivalence''  classes ($m$ and $\delta$ of Eq.~(\ref{quiescence}) 
changing in such a way to preserve the factors $m$$\delta$ and $n$$\delta$). The probability density function $\eta_{Q}(d_{T,\mu,\sigma})$ of the zero-average 
unitary-variance drop diameter sequence has two asymptotic exponential tails: one for the positive and one for the negative values of the rescaled zero-averaged 
diameters (Fig.~\ref{figure4} panels (b) and (c)). A least squares fit  of the exponential tails produce the following values for the decay constants: 
\mbox{$\lambda_{+}$$=$$2$} \mbox{($2$mm$\leq$$d_{T,\mu,\sigma}$$\leq$$6$mm)} and \mbox{$\lambda_{-}$$=$$4.56$} \mbox{($-4$mm$\leq$$d_{T,\mu,\sigma}$$\leq$$-2$mm)}.
\section{Conclusions}
We introduce the concept of quiescence to describe the temporal variability of the rainfall phenomenon. The quiescence captures a fundamental 
relationship (Fig.~\ref{figure2}) between inter drop time intervals, drop diameter and their time ordering. These properties are not detected by the 
flux-like quantities such as rain duration and rain intensity. Using the concept of quiescence, we identify what are the universal properties of the 
rainfall phenomenon. The scaling property of the probability density function of inter drop time intervals during quiescent periods (Fig.~\ref{figure3}) and 
the universal shape for probability density function of drop diameters (Fig.~\ref{figure4}). Our results suggest that the analysis of inter drop time intervals 
and drop diameters sequences and their properties offers a deeper insight than the analysis of the properties of flux-like quantities such as rain duration 
and rain intensity. A comprehensive understanding of the rainfall phenomenon must rest on its drop-like nature.
%
%
\begin{acknowledgements}
M. I. and S. B. thankfully acknowledge Welch Foundation and ARO for financial support through Grant no. B-1577 and no. W911NF-05-1-0205, respectively.
Disdrometer data have been kindly provided by British Atmospheric Data Centre Chilbolton data archive.
We would like to thank Dr. P. Allegrini for his helpful comments and wish all the best to his newborn child. Many thanks to Dr. R. Vezzoli for her help 
and her quick messenger-course on ``the psychology of the feminine gender'': sorry, we failed you. Finally, our eternal gratitude goes to Mr. F. Grosso 
for making us so proud and happy with his beautiful goal at $11$$8^{\prime}$ minute of the second overtime of Italy-Germany (World Cup 2006).
\end{acknowledgements}


\end{document}